\documentclass[preprint,12pt]{aastex}

\begin{document}

\newcommand\tna{\,\tablenotemark{a}}
\newcommand\tnb{\,\tablenotemark{b}}

\title{The Decay of Debris Disks around Solar-Type Stars}
\author{J. M. Sierchio\altaffilmark{1,2}, G. H. Rieke\altaffilmark{1}, K. Y. L. Su\altaffilmark{1}, \& Andras G\'asp\'ar\altaffilmark{1}}
\altaffiltext{1}{ Steward Observatory, University of Arizona, Tucson, AZ 85721}
\altaffiltext{2}{ Currently at Department of Physics, Massachusetts Institute of Technology, Cambridge, Massachusetts 02139, USA}
\email{sierchio@mit.edu}

\begin{abstract}
We present a {\it Spitzer} MIPS study of the decay of debris disk excesses at 24 and 70 $\mu$m for 255 stars of types F4 - K2. We have used multiple tests, including consistency between chromospheric and X-ray activity and placement on the HR diagram, to assign accurate stellar ages. Within this spectral type range, at 24 $\mu$m, $13.6 \pm 2.8 \%$ of the stars younger than 5 Gyr have excesses at the 3$\sigma$ level or more, while none of the older stars do, confirming previous work. At 70 $\mu$m, $22.5 \pm 3.6\%$ of the younger stars have excesses at $ \ge$ 3 $\sigma$ significance, while only $4.7^{+3.7}_{-2.2}$\% of the older stars do. To characterize the far infrared behavior of debris disks more robustly, we double the sample by including stars from the DEBRIS and DUNES surveys. For the F4 - K4  stars in this combined sample, there is only a weak (statistically not significant) trend in the incidence of far infrared excess with spectral type (detected fractions of 21.9$^{+4.8}_{-4.3}\%$, late F; 16.5$^{+3.9}_{-3.3}\%$, G; and 16.9$^{+6.3}_{-5.0}\%$, early K).   Taking this spectral type range together, there is a significant decline between 3 and 4.5 Gyr in the incidence of excesses with fractional luminosities just under $10^{-5}$. There is an indication that the timescale for decay of infrared excesses varies roughly inversely with the fractional lumnosity. This behavior is consistent with theoretical expectations for passive evolution. However, more excesses are detected around the oldest stars than is expected from passive evolution, suggesting that there is late-phase dynamical activity around these stars. 

\end{abstract}

\section{Introduction}
Understanding planetary system formation and evolution is one of the major initiatives in astronomy. Stars form surrounded by protoplanetary disks of primordial gas and dust where planets grow. The material in these disks that does not fall into the star either collects into planets or is dissipated by processes such as photo-evaporation \citep[e.g.,][]{clarke2001} and tidal forces from planets \citep[e.g.,][]{bryden1999}, typically in less than 10 Myr \citep{williams2011}. The evolution of the systems is not complete, as shown by the events that led to the formation of our Moon and to the Late Heavy Bombardment \citep{tera1974}, long after the protoplanetary disk cleared from the Sun. It is very difficult to observe these later stages of evolution directly. However, after dissipation of the protoplanetary disks, a relatively low level of dust production can be sustained through debris produced in planetesimal collisions  to form planetary debris disks \citep{wyatt2008}, which can be detected readily in the infrared over the entire range of stellar lifetime (to 10 Gyr). Debris disks are our best current means of studying planet system evolution over its entire duration. 

About 20\% of the nearby stars harbor debris disks above current detection limits \citep{habing2001,trilling2008,carpenter2009,gaspar2013,eiroa2013}. The likelihood of a detectable debris disk at 24 $\mu$m depends on age, with a higher percentage around young stars than older ones \citep[e.g.,][]{habing1999,rieke2005,gaspar2013}. The expected timescale for evolution of the disk components that dominate in the far infrared indicates that, depending on disk location and density, there might also be a detectable decay in disk incidence at 70 $\mu$m \citep{wyatt2008}. However, it has proven difficult to confirm this prediction definitively \citep{wyatt2008, bryden2006, trilling2008,carpenter2009}. \citet{gaspar2013} have recently demonstrated a drop in infrared excess on the basis of comparison with the predictions of a theoretical model. If such a decay can be confirmed by an alternative analysis, it would help substantially to constrain models and particle properties of debris disks.

In this paper, we explore debris disk evolution in the far infrared (70 - 100 $\mu$m) using a large sample of stars to allow reaching statistically robust conclusions. The paper presents high-quality homogeneous data reductions for a large sample of stars observed with {\it Spitzer}, analyzes their behavior, and then combines this sample with the {\it Herschel}-observed DEBRIS \citep{matthews2010} and DUNES \citep{eiroa2013} samples. We take care in determining stellar ages, since the apparent rate of decay can be strongly influenced by mis-classification of the ages (a small number of young stars mistakenly identified as old ones might dominate the detections among the "old" sample). Section 2 describes the selection of the {\it Spitzer} sample, the determination of stellar ages, and the reduction of the infrared measurements. We determined debris-disk-emitted excesses as described in Section 3. Section 4 presents the analysis of the behavior of these excesses with stellar age, and Section 5 merges this sample and additional {\it Spitzer} data with {\it Herschel} observations to study debris disk evolution in a total sample of about 470 stars. Our conclusions are summarized in Section 6. 

\section{Sample Selection and Data Reduction}

\subsection{Selection Criteria}
\subsubsection{Photometric Criteria}
We used specific criteria to draw samples of stars from the entire {\it Spitzer} Debris Disk Database. All stars were required to have Hipparcos data \citep{vanleeuwen2007}, with parallax errors not to exceed 10\% (the smallest parallax in the sample is 10.7 mas). $V$ magnitudes were taken from Hipparcos, transformed to Johnson $V$ \citep{harmanec1998}. In general the sample is limited to stars with 24 $\mu$m magnitudes less than 7 (i.e., to be brighter than $\sim$ 12 mJy); the largest $V$ magnitude is 8.59. All stars were also required to have observations on the 2MASS $K_S$ system. These measurents were obtained from 2MASS when the measurement errors were indicated to be $<$ 3\%. For stars with saturated measurements, $K$-band photometry from the literature was transformed to the 2MASS system, using relations in \citet{carpenter2001} or \citet{koen2007} or derived for this study. A particularly important set of measurements is from the Johnson Bright Star program \citep{johnson1966a, johnson1966b}, for which transformations are difficult to determine because of dynamic range issues (the faintest stars measured accurately in the infrared in the Bright Star program are too bright to be measured without saturation in 2MASS). In this case, we determined the transformation through intermediate steps to obtain:

\begin{equation}
K_S = K_{Johnson} - 0.0567 +0.056(J-K)_{Johnson}
\label{equ1}
\end{equation}

\noindent
When at least three measurements had been made of a star in the Bright Star program \citep{johnson1966a}, we found that the transformed result was within our 3\% accuracy requirement, as judged by the scatter around standard colors.  In the following discussion the $V$ and $K$ magnitudes will refer consistently to $V$ magnitudes transformed to the Arizona (Johnson) system from Hipparcos measurements and to $K_S$ magnitudes on the 2MASS system, either from 2MASS or transformed to that system.

Additional measurements were obtained at SAAO on the 0.75-m telescope using the MarkII Infrared Photometer (transformed as described by \citet{koen2007}), and at the Steward Observatory 61-Inch with a NICMOS2-based camera with a 2MASS filter set and a neutral density filter to avoid saturation. These measurements will be described in a forthcoming paper (Su et al. in preparation).

 Our study is confined to solar-like stars, defined as having spectral types in the range of F4-K2 and of luminosity classes IV-V (obtained from SIMBAD). This places the Sun as G2 in the middle of our spectral type range. To guard against the uncertainties in spectral classifications, we used the photometric colors to select stars with $1.05 < V-K < 2.0$, which corresponds to the desired range of spectral types \citep{tokunaga2000}. The spectral types of the stars retained were consistent with their colors, although the color selection eliminated a few stars with indicated spectral types inside the desired range.

\subsubsection{Other parameters}

The original programs in which our sample stars were measured are identified in Table 1. A large majority (93\%) come from seven {\it Spitzer} programs: 1.) the MIPS GTO sun-like star observations \citep{trilling2008}; 2.) Formation and Evolution of Planetary Systems (FEPS) \citep{meyer2006}; 3.) Completing the Census of Debris Disks \citep{koerner2010}; 4.) potential SIM/TPF targets \citep{beichman2006}; 5.) an unbiased sample of F-stars \citep{trilling2008}; and 6.) two coordinated programs selecting stars on the basis of indicators of youth \citep{low2005,plavchan2009}. All of these programs were based on unbiased samples selected in differing ways. The remaining 7\% of the sample come from a broad range of programs, but, with the single exception of HD 206893, none made selections on the basis of prior detections. Another source of bias is sources left off a selection because they are committed to guaranteed programs. This issue is most important for the MIPS solar type program, and \citet{bryden2006} list four stars that would otherwise have been in their sample. Of them, HD 10647 is in our sample and HD 17206 is in the DEBRIS sample and will be included when we merge with it. HD 10700 \citep{chen2006} does not appear to have an excess at the wavelengths of interest (so it would have little influence on our results), while $\epsilon$ Eri is not in our sample. To first order, it and HD 206893 offset biases, since they are both young stars with substantial excesses. However, these details demonstrate that any sample compiled from a collection of other samples defined on different bases are subject to selection biases.

We checked SIMBAD and the literature for indications of stars being members of multiple star systems. Such stars were only allowed in the final sample if they had a separation of less than 1 arcsecond or greater than 12 arcseconds. For typical distances for the sample members ($\ge$ 8 pc), 12 arcsec corresponds to a separation $\ge$ 100 AU and hence avoids the possible bias against debris disks for binary separations less than this value \citep{trilling2007, rodriguez2012}. Close binaries were allowed because they do not appear to suppress outer debris disks (emitting in the far infrared) \citep{trilling2007}. The original samples from which ours is drawn were not selected on the basis of rotational characteristics, although on the basis of gyrochronology we expect systematically higher {\it vsini} for the younger stars. After assembly of a preliminary sample, we carefully screened each star repeatedly for issues such as blending or complex backgrounds. The stars rejected in these tests were biased slightly toward being young but the distribution did not differ significantly from the age distribution of the entire sample (rejecting 10 in the 0 - 750 Myr range, 4 in 750 Myr - 5 Gyr, 4 in $>$ 5 Gyr). Finally, we consider whether metallicity could bias our results, given the dependence of detected planets on metallicity \citep{fischer2005, johnson2010}. However, a number of studies imply that the dependence of debris-disk excesses on [Fe/H] is weak \citep{trilling2008, maldonado2012}, and that the correlation between the presence of planets and debris disks, although present, is also weak \citep{bryden2009,maldonado2012}. Any bias due to metallicity effects can also be discounted because the average [Fe/H] of the stars of ages $<$5 Gyr is -0.13, whereas for those at ages $>$ 5 Gyr, it is virtually identical at -0.15.

Given the requirements on spectral type and apparent magnitudes, the sample is almost entirely (95\%) within 50 pc of the sun. The stars beyond this distance are representative of the sample as a whole: in average metallicity, [Fe/H] = -0.14; in average spectral type, F9V; in age, seven 0 - 700 Myr, three 700 Myr - 5 Gyr, three $>$ 5 Gyr; and in incidence of excesses, 30\%.

\subsubsection{Ages}
Accurately-measured ages are essential to determine the decay of debris excesses. If the excesses decay substantially over the age range of the sample of stars, a few young stars incorrectly classified as being old can seriously affect the apparent rate of decay. To ensure the accuracy of our results, we took a conservative approach by using three different methods to estimate ages: (1) placement on the HR-diagram, (2) chromospheric activity indices (CAI), and (3) X-ray luminosity. For the latter two indicators, we used the calibration of Mamajek \& Hillenbrand (2008).

Each of these methods has weaknesses, making it desirable to use all three to verify the consistency of the results. Traditionally, use of the HR diagram can lead to relatively large errors in ages; the method is intrinsically inaccurate for stars near the main sequence, and it can also give incorrect results for reasons such as undetected stellar multiplicity. However, used with care the HR diagram is useful for determining ages of stars old enough to have left the main sequence. In comparison, CAI ages are well-calibrated for stars older than $\sim$ 300 Myr and younger than $\sim$ 4 Gyr (Mamajek \& Hillenbrand 2008). For stars younger than 300 Myr, the issues are complex and will be discussed below. For stars older than 4 Gyr, the calibration of Mamajek \& Hillenbrand (2008) rests on only 23 stars with accurate ages from placement on the HR diagram. X-ray detections are also useful for young stars, but measurements of old stars are relatively uncommon. Because of their differing strengths and weaknesses, we tested the consistency of the indicators before proceeding with their use. For these tests, we drew stars from three samples studied extensively for debris disks: our combined {\it Spitzer} sample and those for the Herschel key programs DUNES and DEBRIS. 

We compare the ages obtained via chromospheric activity with those from X-ray luminosity in Figure 1. The sources for the CAI, log$R'_{HK}$, can be found in Table 2 and in \citet{gaspar2013}. X-ray detections were taken from the ROSAT data \citep{voges1999, voges2000}, with the requirement of positional matches within 30 arcsec and - 0.8 $<$ HR1 $<$ 0; this constraint on the hardness ratio (HR1) was adopted to avoid AGN (HR1 $>$ 0.5), stars that are possibly flaring (0 $<$ HR1 $<$ 0.5) and background thermal sources (HR1 $<$ -0.8) (T. Fleming, private communication). The trend line in Figure 1 (fitted by linear least squares with equal weight for all points) has a nominal slope of 1.28 $\pm$ 0.05, with a tendency for the points for the oldest stars to have larger ages from the CAI than from X-ray luminosity. This tendency could arise from a selection effect (Eddington bias). Only a minority of old stars are detected by ROSAT, and the limited ratios of signal to noise mean that the detected sample will tend to include stars slightly overestimated in X-ray flux (due to statistics), leading to underestimated ages. 

We tested this possibility on the stars in our sample with ages $\ge$ 5Gyr based on the CAI. We used the CAI age estimates, the parallaxes, and photometric data to predict the ROSAT count rate for each star, which we compared with the ROSAT data. We evaluated the detection limits for each star from typical measurements for faint sources in the ROSAT Faint Source Catalog (Voges et al. 2000) in a field of 1 degree radius around the star. These detection limits and the predicted counts for the star were used to calculate the probability that it would have been above the detection threshold (Voges et al. 2000) for the survey. The sum of these probabilities over the entire sample of stars was 18 and is an estimate of how many detections are expected if the CAI ages agree perfectly with those from X-ray emission. In fact, 27 stars could be identified with cataloged X-ray sources (Voges et al. 1999, 2000); however, three of them were so discordant from the count rate predictions for their ages and distances that the ROSAT detections are suspect and may arise from background sources. Therefore, we predict 18 detections and find 24, indicating that the X-ray fluxes are slightly brighter than expected from the Mamajek and Hillenbrand (2008) calibration. However, at the 1 $\sigma$ level, virtually no correction is indicated; in addition X-ray emission and the CAI are measures of closely related stellar parameters. We therefore used the Mamajek and Hillenbrand (2008) calibration "as-is." Nonetheless, it would be desirable to test the X-ray age calibration against a larger sample of old stars. 

We next used the {\it Spitzer}-DEBRIS-DUNES sample of stars (our sample plus those in \citet{gaspar2013}) to compare CAI ages with those from the HR diagram. We select $V-K$ and $M_K$ for the axes of the HR diagram, since the long wavelength baseline of $V-K$ makes it an accurate indicator of stellar temperature \citep{masana2006} and the K-band should be a relatively robust luminosity indicator, since it is in the Rayleigh-Jeans spectral regime for our sample of stars. To place the stars on the HR-diagram (Figure 2), we first eliminated from this expanded sample all cases with detected close companions (within 12$"$) that would be included in or confuse the photometry of the star. The $V-K$ color is quite sensitive to metallicity \citep{li2007}; $M_V$ is also metallicity sensitive but this effect should be reduced for $M_K$ given the simpler spectrum across the $K$ window. To calibrate the effects of metallicity on the HR diagram, we determined a linear transformation that made the empirical metallicity-dependent isochrones of \citet{an2007} coincide over the range -0.3 $\le$ [Fe/H] $<$ 0.2. We then reversed this transformation and applied it to each star so that stars of differing metallicity can be compared on a single HR diagram: 

\begin{equation}
corrected ~~ M_{K_{S}} = M_{K_{S}} + 0.5[\frac{Fe}{H}]
\label{equ2}
\end{equation}
\begin{equation}
corrected ~~ V-K_S = V-K_S - 0.25[\frac{Fe}{H}]
\label{equ3}
\end{equation}

\noindent
Metallicity measurements were obtained from a number of sources \citep[e.g.,][]{marsakov1995, gray2003, valenti2005, gray2006, ammons2006, holmberg2009, soubiran2010}; when more than one measurement was available, they were averaged. Stars with [Fe/H] $<$ -0.4 were eliminated in this comparison (but not in our final study if we had reliable age indicators) because our metallicity correction is not calibrated for them. We used the isochrones from the Padua group for the final comparisons \citep{bonatto2004, marigo2008}, because they have better coverage for older stars than those of \citet{an2007}. An empirical correction of -0.14 magnitudes was made to the absolute $K_S$ magnitude on the isochrones to match the measured values for the main sequence; this correction removes systematic issues in matching the observational and theoretical photometric systems. 

The majority of stars with ages indicated from the CAI greater than 6 Gyr fall between the 6 and 11 Gyr isochrones, as shown in Figure 2. Figure 3 compares the ages; a few stars were eliminated from the sample based on this comparison because their ages from the HR diagram were much younger than from the CAI. Since it is difficult for binarity or other issues to produce such discrepancies, the CAI ages for these stars are suspect. There are 38 stars in the sample in Figure 3 older than 4 Gyr; the trend line (fitted by linear least squares with equal weight for all points) has a slope of 0.985 $\pm$ 0.027, insignificantly different from 1, and there is no apparent deviation from the trend line with increasing age. We confirm that the CAI/X-ray calibration of Mamajek \& Hillenbrand (2008) is consistent with isochrone ages for stars older than 4 Gyr.

With this validation, we have used chromospheric activity and X-ray luminosity as our primary age indicators. We supplemented these indicators with $logg < 4$, which for stars over the relevant spectral range (F4 - K2) shows that a star is above the main sequence, and with gyrochronology ages when available. We then checked that the age was consistent with the placement of the star on the HR diagram. Stars where the initial criteria indicated ages $>$ 5 Gyr,. but which fell in an inappropriate region on the HR diagram, were not used further. The final {\it Spitzer} sample is presented in Table 2 with the $V$ and $K_S$ measurements, ages, and Spitzer measurements at 24 $\mu$m and 70 $\mu$m. The age distribution of the sample is shown in Figure 4; there are 255 stars with age estimates, of which 238 have useful 70 $\mu$m measurements.  

We now discuss the uncertainties in the ages. Taking ages directly from \citet{mamajek2008} would be circular, since we use the same calibration and some of the same data sources. However, comparing with those ages is revealing; there is a scatter of 0.15 dex. These differences primarily arise because of data published since the publication of \citet{mamajek2008}, and they illustrate that there can be significant uncertainties despite the care we have taken in estimating stellar ages. Another issue is that there is a large spread in chromospheric activity around 50 - 300 Myr \citep{mamajek2008}. As a result, a star with a low level of activity may still be within this age range. Based on Figure 5 of \citet{mamajek2008}, it is possible that about 20\% of the stars in the 50 - 300 Myr range are incorrectly placed in the 0.6 - 2.5 Gyr one if the age is assigned purely on the basis of the CAI. This rate of incorrect age assignment is an upper limit because we use multiple indicators to check against each other.  More importantly, we find that the incidence of debris disk excesses does not vary rapidly across the 0.1 - 2.5 Gyr age range (see Section 5.2), so a modest number of age errors within this range will have little effect on our results.

\subsection{Data Reduction}

\subsubsection{24 $\mu$m}
We re-processed all the mid- and far-infrared data for these stars as part of a ~{\it Spitzer}~ legacy catalog, using the MIPS instrument team Data Analysis Tool \citep{gordon2005} for the initial reduction steps. In addition, a second flat field constructed from the 24 $\mu$m data itself was applied to all the 24 $\mu$m results to remove scattered-light gradients and dark latency \citep{engelbracht2007}. The processed data were then combined using the World Coordinate System information to produce final mosaics with pixels half the size of the physical pixel scale (which is 2.49$"$). We extracted the photometry using point-spread function (PSF) fitting. The input PSFs were constructed using isolated calibration stars and have been tested to ensure that the photometry results are consistent with the MIPS calibration \citep{engelbracht2007}. Aperture photometry was also performed, but the results were only used as a reference to screen targets that might have contamination from nearby sources or background nebulosity.

The random photometry errors were estimated based on the pixel-to-pixel variation within a 2 arcmin square box centered on the source position.  The measured flux densities (using 7.17 Jy as the zero magnitude flux; \citep{rieke2008}), and associated random errors are listed in Table 2. The errors from the photometry repeatability ($\sim$ 1\% at 24 $\mu$m; \citet{engelbracht2007}) are included, which were measured on stars of similar brightness to our sample members (and hence give an estimate of photon noise as well as other systematic photometric errors). For the stars with indicated 24 $\mu$m excesses, we have compared with the WISE W3 - W4 color. In all cases where the WISE W4 measurement had adequate signal-to-noise, the MIPS measurement was confirmed\footnote{The WISE data for HD 25457 show multiple sources and appear to be confused; the MIPS data are valid.}. At least two of the stars in our sample have strong excesses at 24 $\mu$m and weak ones at 70 $\mu$m, placing them in the rare category of predominantly warm debris disks (HD 1466, for which the weak far infrared excess is confirmed with Herschel \citep{donaldson2012} and HD 13246 \citep{moor2009}). 

\subsubsection{70 $\mu$m}

The 70 $\mu$m observations were obtained in the default pixel scale (9.85$"$). After initial reprocessing, artifacts were removed from the individual 70 $\mu$m exposures as described in \citet{gordon2007}. The exposures were then combined with pixels half the size of the physical pixel scale. The image of each source was examined to be sure that it was free of confusing objects such as IR cirrus emission or background sources. A number of objects were rejected at this stage. PSF photometry was used to extract flux measurements for the 'clean' sources, utilizing STinyTim PSFs that had been smoothed to match the observations (accounting primarily for the effects of the pixel sampling) and with the position fixed to that found at 24 $\mu$m. The results were calibrated as in \citet{gordon2007}. The random photometry errors were based on pixel-to-pixel variations in a 2 arcmin field surrounding the source and are tabulated (including a 5\% allowance for photometric non-repeatability \citep{gordon2007}) along with the flux densities in Table 2. 

\section{Determination of Excesses}
\subsection{24 $\mu$m Excesses}
Excesses at 24 $\mu$m were detected by using a V-K vs. K-[24] plot (Figure 5). The locus of stars without excesses was determined from a total of $\sim$ 1320 stars drawn from the {\it Spitzer} archive, reduced identically to the stars in our sample, and with K-magnitudes from the same sources. The resulting fit is \citep{urban2012}:
\begin{equation}
\resizebox{.9\hsize}{!}{$K_S-[24] = 0.0055 - 0.0134x + 0.06555x^{2} - 0.1095x^{3} + 0.066415x^{4} - 0.01458x^{5} + 0.001101x^{6} - 0.000003x^{7}$}
\label{equ4}
\end{equation}
\noindent
for x $>$ 0 where x is $V-K_S-0.8$. The scatter around this fit was determined for all 1320 stars by fitting a Gaussian to the residuals. Because a significant number of stars have excesses, the fitting parameters need to be selected to avoid a bias. We first used outlier rejection based on Huber's Psi Function, set just to avoid rejecting any points below the photospheric value. We used the Anderson-Darling (A-D) test to confirm that the remaining distribution around the photospheric values was consistent with a normal distribution. Inspection of the results indicated that there were a number of stars with excesses (the positive side of the distribution above 0.03 in residual had 127 stars compared with 96 for the negative side of the distribution below -0.03), so before calculating the A-D p-value, we replaced the measures above 0.03 with the inverse of the ones below -0.03. The A-D test then indicated that the distribution around the photospheric value was indistinguishable from a Gaussian (p = 0.14). We therefore fitted a Gaussian to these points, finding a standard deviation of 0.0304.

To provide a more intuitive check of this approach, we carried out a second analysis where we fitted a Gaussian excluding values more than 2-$\sigma$ above the photospheric trend line (that is above 0.06). As with use of the Psi Function, this strategy makes the fit independent of infrared excesses detected in some of the stars of the sample but is more subjective than the outlier rejection approach.  A demonstration of this result can be found in \citet{ballering2013}, which demonstrates the excellent fit of a Gaussian around the photospheric value as well as showing the population of excesses. In the current case, this Gaussian indicated a standard deviation of 0.0295. That is, the two Gaussian fits to the peak of the measurement distribution around the photospheric value agree closely in width. We adopt a value of $\sigma$ = 0.03. Therefore, any star with a $K_S-[24]$ $>$ 0.09 magnitudes larger than predicted for its photosphere is considered to have an excess at 24 $\mu$m (nominally, the number of false identifications in our sample is then expected to be less than 0.5). None of the stars older than 1000 Myr have excesses based on this requirement; 23 of the younger stars do, a fraction of 0.136$ \pm 0.028$ (errors from binomial statistics). Stars with excesses are listed in Table 3. 

\subsection{70 $\mu$m Excesses}
A method similar to that described in \citet{trilling2008} was used to determine which stars had excesses at 70 $\mu$m. We computed an excess ratio, R\footnote{R is defined as the stellar flux density divided by that expected from the photosphere alone, R = $F_{70}/P_{70}$.}, for each star by predicting a photospheric flux density based on the measured flux density at 24 $\mu$m divided by 9.05 (the nominal ratio of flux densities between 24 and 70 $\mu$m) and then divided by the ratio of the measurement to the predicted photospheric value at 24 $\mu$m if this ratio exceeded one. We used the figure of merit:
\begin{equation}
\chi_{70} = (F_{70} - P_{70}) / \sigma
\label{equ5}
\end{equation}
\noindent
to quantify the significance of an excess, where $F_{70}$ is the measured flux density at 70 $\mu$m, $P_{70}$ is the predicted photospheric flux density, and $\sigma$ is the uncertainty in the flux density measurement at 70 $\mu$m (the uncertainty in $P_{70}$ is negligible compared with that in $F_{70}$ for our entire sample).
In total 38 stares younger than 5 Gyr and 4 stars older than this age are considered to have excesses, as illustrated in Figure 6 and listed in Table 3. For the young sample, the probability of having an excess (at 70 $\mu$m) is 38/169 = 0.225$ \pm 0.036$. Similarly for the old sample, the probability of an excess is 4/86 = 0.047$^{+.037}_{-.022}$ (all errors are calculated from the "exact" or Clopper-Pearson method, based on binomial statistics).  Three of the four detections around old stars are at $\chi_{70}$ $>$ 4.5 and should be very reliable (see Figure 6). These three systems are similar in amount of excess to typical younger stars; the ratios of the fluxes at 70 $\mu$m to the photospheric fluxes are close to the median for our entire sample.  

\subsection{Stars Older Then 2.5 Gyr with 70 $\mu$m Excesses}

The major conclusion from this paper will rest on the small number of old stars with excesses. The reliability of this number is influenced both by the reliability of the identification of excesses, and by the accuracy of our age estimates. With regard to the reliability of the excesses, we have compared with the results from the Herschel DUNES and DEBRIS surveys (as summarized in \citet{gaspar2013} and \citet{eiroa2013}). There are 50 stars in our sample also in the Herschel samples. Using the same 3-$\sigma$ selection criterion, 41 of the 50 stars show no evidence for excess emission with either telescope\footnote{HD 1237, 4308, 4614, 10307, 17051, 20630, 23754, 43162, 50692, 55575, 62613, 76943, 78366, 79028, 84737, 86728, 89269, 90839, 99491, 101501, 102365, 102438, 109358, 114710, 115383, 126053, 126660, 130948, 142860, 143761, 145675, 160691, 162003, 168151, 180161, 189245, 190422, 202275, 210277, 210302, and 212330}. Of the remaining nine stars, six show significant excesses in the data from both telescopes, (HD 20794, 30495, 33262, 76151, 199260, and 207129) and one more (HD 153597) has an excess close to 3 $\sigma$ in both sets of data. The remaining two stars are HD 30652, and 117176. The first case is just over the 3-$\sigma$ threshold with PACS and at 1 $\sigma$ excess with MIPS (assuming 5\% photometric errors plus the statistical errors); the two values agree within 1.8 standard deviations. However, the weighted average of the two measurements has a net excess at less than 3-$\sigma$ significance, so we classify it as a non-excess star. For HD 117176, the MIPS measurement (at 70 $\mu$m) is right on the photospheric value while the PACS one (at 100 $\mu$m) is well above it; the two differ nominally by 4 $\sigma$. The implied very cold spectral energy distribution for the excess suggests that it could arise from a background galaxy \citep{gaspar2014}, so we have listed this star as not having an excess.

Because of the residual uncertainties in age, we will discuss the stars with excesses and wth ages indicated to be greater than 2.5 Gyr (since we will show that the incidence of excesses drops substantially between 3 and 5 Gyr). At the same time we tabulate their spectral types to probe whether they are characteristic of the full sample. We begin with the stars indicated to have ages between 2.5 and 5 Gyr: {\it HD 20807}, G0V: The placement on the HR diagram indicates that this star is less than 3.5 Gyr old. As indicated in Table 1, multiple sources indicate chromospheric activity levels too high for the star to be 5 Gyr or older. {\it HD 33636}, G0V: The placement on the HR diagram indicates an age $<$ 3.5 Gyr. \citet{isaacson2010} report 17 measurements of chromospheric activity that are, with a single exception, consistent with the quoted age; the age should be robust against measurement errors and variations in activity. {\it HD 207129}, G2V: There are suggestions that we have overestimated the age of this star (\citep{montes2001, mamajek2008, tetzlaff2011} but no indication that it could be older than 5 Gyr. We now discuss the stars older than 5 Gyr and with 70 $\mu$m excesses: {\it HD 1461}, G0V: Two measurements \citep{wright2004, gray2003} agree that the chromospheric activity of this star is very low, corresponding to an age of 7 to 8 Gyr. The placement on the HR diagram indicates a lower limit of 6 Gyr.  {\it HD 20794}, G8V: Three measurements \citep{henry1996, gray2006, schroeder2009} agree that this star has low chromospheric activity, indicating an age of $\sim$ 7 Gyr. The position of the star on the HR diagram is inconclusive with regard to age. {\it HD 38529}, G4V: Multiple measurements \citep{wright2004, white2007, schroeder2009, isaacson2010, kasova2011} agree on the low chromospheric activity of this star, corresponding to an age of about 6 Gyr. The value of the CAI reported by \citet{isaacson2010} is based on more than 100 measurements. The surface gravity also places this star well off the main sequence. {\it HD 114613}, G3-4 V: Four measurements \citep{henry1996, jenkins2006, gray2006, isaacson2010} agree that this star has a very low level of chromospheric activity, corresponding to an age of about 8 Gyr. The value from \citet{isaacson2010} is based on 28 individual measurements. The placement of the star on the HR diagram indicates an age $>$ 5 Gyr; it also has low surface gravity, consistent with its position well above the main sequence.

\section{Analysis}

In this section, we first show that the rate of excess detection is independent of spectral type, confirming the result from \citet{gaspar2013}. As a result, to obtain good statistics, we can consider the evolution of the debris excesses combining the results for all spectral types without loss of information. We then analyze the time evolution of the excesses for the {\it Spitzer} sample. In the following section, we first show that the stars measured in the DEBRIS and DUNES programs behave consistently with the {\it Spitzer} sample and then combine all the data to double the number of stars. This larger sample shows that the incidence of far-infrared excesses evolves substantially with stellar age (see also \citet{gaspar2013}).

\subsection{Detection Statistics}

Because our samples span late F-, G-, and early K- spectral types, we can report excess fractions for each type. To compute these fractions, we have consider only stars less than 5 Gyr old, since stellar evolution would otherwise be an issue for the F stars. Similarly, we consider only excesses at 70 $\mu$m, since the 24 $\mu$m excesses evolve over an age range of a Gyr. We find fractions with excesses of 16/63, 20/98, and 2/8 respectively for F4 - F9, G, and K stars. We can combine our results with those from DEBRIS and DUNES \citep{gaspar2013, eiroa2013}, using the same definition for detection of an excess ($\chi > 3$). We have omitted HIP 171, 29271, and 49908 from this combined sample because their 160 $\mu$m excesses are likely to arise through confusion with background galaxies \citep{gaspar2014}; we have also omitted HIP 40843, 82860, and 98959 because the reported excesses at 100 $\mu$m \citep{gaspar2013} may also be contaminated \citep{eiroa2013}. The net results are 23/105, 22/133, and 10/69 respectively for F4 - F9, G, and K stars. The corresponding fractions are 0.219$^{+0.048}_{-0.043}$, 0.165$^{+0.039}_{-0.033}$, and 0.145$^{+0.055}_{-0.043}$. If the ten stars later than K4 are excluded (all from the DEBRIS and DUNES samples and none of which have excesses), the fraction for K stars rises to 0.169$^{+0.063}_{-0.050}$; the errors are 1 $\sigma$, computed using the "exact" or Clopper-Pearson method, based on the binomial theorem. Although there may be a weak trend with spectral type, the incidence of excesses is (within the errors) the same for late F through early K-type stars. This conclusion is strengthened if we take account of the possible bias of including HD 206893 (F5V) and excluding $\epsilon$ Eri (K2V) in our sample, as discussed in Section 2.1.2. The results of previous studies are mixed, but our result is in general agreement with \citet{trilling2008} who reported no clear trend of infrared excess detection with stellar type.

\subsection{70 $\mu$m Excess Evolution: K-M Test}

Previous studies have attempted to quantify the decay rate for debris disks by one of two different approaches \citep{rieke2005, trilling2008}: counting the number of stars with detected excesses (i.e., basing the statistics on small excesses) or fitting the largest excesses (i.e., matching the apparent upper envelope of the excesses). Neither of these approaches takes rigorous account of all the data and each is subject to selection effects. In the first case, if the achieved sensitivity is not adequate to detect the stellar photospheres uniformly, the minimum detectable excess is not uniform, whereas in the second case the upper envelope of the excesses is ill-defined and based on small number statistics. Ultimately, these studies found either no or only weak evidence of disk evolution with age. The underlying problem can be described in terms of data censoring, that is, the non-uniform detection limits result in target stars being removed from the study at the levels of excess represented by upper limits.

A standard statistical approach to such a situation is to use the Kaplan-Meier (K-M) estimator, since it naturally allows for censored data \citep{feigelson2012}. We have used the version of the K-M estimator provided by \citet{lowry2013}, which bases its confidence level estimates on the efficient-score method \citep{newcombe1998}. We divide the sample into three equal parts corresponding to the age ranges $<$ 750 Myr, 750 - 5000 Myr, and $>$ 5000 Myr. Based on the detection limits and  identifications of excesses in Table 2, we find that the probable numbers of excess ratios R $>$ 1.5 at 70 $\mu$m are 27\% with 95\% confidence limits of 18\%  and 38\% for stars of age $<$ 750 Myr and and 22\% with confidence limits of 14\% and 33\% for those between 750 and 5000 Myr. That is, the K-M test does not identify a significant change for stars younger than 5 Gyr. The overall probability for stars $<$ 5 Gyr old is 24\% with 95\% confidence limits of 18\% and 34\%, while the corresponding values for stars $>$ 5 Gyr in age are 5.3\% with 95\% confidence limits of 1.8\% and 14\%.The drop in incidence of excesses is a factor of 4.5, at greater than 2$\sigma$ significance. Our discussion in Section 3.3 demonstrates that errors in age or in identification of excesses are unlikely to change significantly the result that there are only four excesses in the portion of our sample older than 5 Gyr. 

\subsection{A Test and Confirmation}

In principle, the Kaplan-Meier estimator works under the assumption of "random" censoring, i.e., that the probability of a measurement being censored does not depend on the source property being probed, in this case stellar age. We deal with debris disk measurements in terms of their significance level (e.g., $\chi_{70}$) to allow for differing depths relative to the photosphere; about half of the sample is measured deeply enough ($\ge 3\sigma$, or $\ge 3 \sigma$ on the expected photospheric level) to be detected. Thus the random censoring assumption may not be completely satisfied. For comparison, we therefore will test for the evolution of 70 $\mu$m excesses using a different method based on the shape of the probability distribution of excesses, which is also able to account for non-uniform detection limits (a similar approach has been used by \citet{gaspar2013}). This method will also confirm that the underlying assumption for the K-M test - that the censoring is independent of age - is satisfied for our sample.

We begin by assuming that a sample of stars will have some probability distribution of excess ratios. Integrating this distribution upward from the detection threshold gives the probability that a star has a detectable excess. Lower limits for the integrals are given by:

\begin{equation}
LL = 1 + 3(\sigma_{70}/P_{70})
\label{equ6}
\end{equation}

\noindent
where 1 represents the expected flux ratio of detected to photospheric flux density for no excess, $\sigma$ is the uncertainty on detected flux density, $F_{70}$, and $P_{70}$ is the predicted photospheric flux density. The factor of three represents the 3$\sigma$ threshold placed on $\chi$ to have a significant excess. For each star in a sample, the integral of the probability distribution from this lower limit to infinity gives the probability of detecting an excess. The values for all the stars can be summed to obtain the total number of predicted excesses in the sample. The probability distribution can be refined by adopting a number of detection thresholds and comparing the sums over the sample stars with the number of detected excesses, testing the probability distribution as a function of amount of excess. Once the final model is obtained, it can be treated as a parent distribution for comparisons with other samples of stars. That is, under the null hypothesis that a sample is identical in excess behavior to the one for which the probability distribution was determined, we can use the distribution to compute how many detections should be achieved in the second sample and compare this prediction with the number of actual detections. 

To apply this approach, we evaluated several possibilities (Gaussian, lognormal, power law, and exponential) for the shape of the distribution of excess ratios for the sample of stars younger than 5 Gyr and with signal-to-noise ratios at 70 $\mu$m of at least 2 (we fitted below the $\chi$ = 3 reliable detection limit to be sure the distribution was well-behaved). We refined initial guesses for the fits by computing the lower integration limit for each star, regardless of whether it was detected, as in Equation (6). We defined the upper limit for the integral to be a large enough that the value had converged. We required the sum to be 38, the total number of stars with indicated excesses ($\chi > 3$). We varied the width-parameter and the normalization constant, leaving the sum unchanged (for the Gaussian and lognormal models, we only varied the width-parameter $\sigma$ and calculated the appropriate normalization factor; for the exponential and power law distributions, the exponential factor was similarly varied - it is the only other free parameter for these models). We then used least-squares minimization to fit the integral sums for each function at selected excess thresholds to the number of actual detections above that threshold. We chose 10, 20, 30, 50, 60, 90, 100, and 200, as the test thresholds. The results of these tests are given in Table 4. The lognormal and power law shapes produced the best fits.
We found that the lognormal function was very consistent with the distribution of lower limits in the sample, while the power law was not. We therefore conclude that the excess ratios and lower limits are best fit with a lognormal shape: 

\begin{equation}
\frac{C}{x\sqrt{2\pi}\sigma}e^{-\frac{(\log{x} - \mu)^2}{2\sigma^2}}
\label{equ7}
\end{equation}

\noindent
This result is consistent with previous work \citep{andrews2005,wyatt2007a,kains2011,gaspar2013} that also found that the distributions of young disk masses or excesses could be fitted with lognormal functions. To test our model, we divided the sample into two age sub-bins of 80 and 79 stars: age $<$ and $\ge$ 750 Myr. Within these sub-bins, 21 and 18 stars with excesses were detected, respectively. The model predicts 18.2 and 19.8 excesses for those bins, confirming the fit.

Using lower integration limits computed from the data for the stars older than 5 Gyr, the lognormal model predicts 21.6 excesses at 70 $\mu$m, assuming no decay relative to the young stars. Thus, there is no significant dependence on the predicted number of excesses with stellar age; the three age divisions respectively have 18.2, 19.8, and 21.6 predicted detections, from virtually identical numbers of stars in each bin. This confirms the underlying assumption in the K-M test that the probability of censoring is not a function of stellar age.

We can also use this approach to confirm the result from the K-M test.  Since only four excesses were detected, we can take the ratio of the predicted number to the detected number to obtain an estimated decay by a factor of $\sim$ 5. The statistical significance of this result is dominated by the poor statistics for the old star sample; it indicates a decay by a factor greater than two (2 $\sigma$ limit). Our modeling therefore establishes independently that debris disk excesses decay at 70 $\mu$m over time scales of 5 Gyr.  Furthermore, our division of the young sample into two parts (younger and older than 750 Myr) and the finding of an equal incidence of excess in them indicates that the decay occurs late in the evolution of the debris disks, past 1 Gyr. 

To place constaints on the accuracy of the model, we used a Monte Carlo approach. We created random samples of the same size as our young ($<$ 5 Gyr) sample, using the same parameters as those in the original probability distribution. For each random sample, we recomputed the best-fit parameters. We also fit the lower limits described above to a lognormal distribution and used that distribution to generate random samples of lower limits.  The recomputed fits for the random star samples and random lower limits were used to compute a predicted number of excesses for the generated random sample of stars. Over 10000 trials, the average number of excesses predicted for the young sample was 38.0 with a variance of 5.0. We repeated the procedure for the old sample by fitting the old sample lower limits to their own lognormal distribution, generating random old sample lower limits, and recomputing the fit parameters. We still used the recomputed best-fit parameters for the young star sample, and again calculated the predicted number of excesses in the old sample. The average number of excesses predicted for the random old samples was 21.6 with a variance of 1.3, over 10000 trials. The error in the predicted number of excesses is therefore dominated by the statistical uncertainty in the number of detected young stars. Combining the uncertainties quadratically, the predicted number of excesses is 21.6 $\pm$ 2.5, confirming our 2-sigma limit is valid.

This result is in excellent agreement with the conclusions from the K-M estimator. The consistency of the two different methods gives confidence that the conclusions drawn from the K-M estimator are correct.

\section{More detailed evolution of debris disks}

Another issue with previous studies (e.g., \citet{bryden2006, carpenter2009} is small samples and as a result poor statistical significance; this also limits our conclusions from the {\it Sptizer} sample. To examine the evolution of debris disks with better statistical weight, we combine the sample described in this paper with that from DUNES \citep{eiroa2013} and DEBRIS \citep{matthews2010}. We take the ages of these stars from \citet{gaspar2013}, discarding all of quality "1"; these ages were estimated as in this paper except for omission of the step to reject cases where the HR diagram and CAI were not in agreement  (we therefore added a check on the HR diagram as with the {\it Spitzer} sample for stars older than 5 Gyr).  We also take the {\it Spitzer}/MIPS 70 $\mu$m and DEBRIS 100 $\mu$m photometry from \citet{gaspar2013}, with the DUNES photometry from \citet{eiroa2013}. We do not consider disks possibly detected only at 160 $\mu$m (see \citet{gaspar2014}), and also discard photometry listed by \citet{eiroa2013} as likely to be confused with IR cirrus or background galaxies. We merged the measurements into a single indicator of an excess at 70 - 100 $\mu$m, and determined the significance of each excess as a $\chi$ value analogous to the definition in equation (5) (see \citet{gaspar2013}). 

\subsection{Behavior of DEBRIS and DUNES Samples}

The number of stars in the combined DEBRIS and DUNES samples within the relevant range of spectral types and with valid ages is 231, nearly as large as the purely {\it Spitzer} one. We have therefore first analyzed it independently as a check of the results in the preceeding section. In this case, to divide the sample into thirds, we adopt age ranges of 0 - 1500 Myr, 1500 - 5000 Myr, and $>$ 5000 Myr. For excesses with R $>$ 1.5, we find the detection probabilities are respectively 17\% (95\% confidence limits of 10\% and 28\%),  19\% (limits of 11\% and 19\%) and 11\% (limits of 5\% and 21\%). These values are in agreement with those from the {\it Spitzer} sample, although a bit lower and hence the drop in the incidence of excesses in the oldest age range is also indicated but at a slightly less statistically significant level (slightly less than 2$\sigma$). The overall similarity of the two samples is not surprising, given that they include virtually identical numbers of stars and that the numbers of infrared excesses detectable by {\it Spitzer}/MIPS and {\it Herschel}/PACS are also similar (see \citet{gaspar2013}).

Again, this conclusion rests on the reliability of the assigned ages, so we review those for the old stars (i.e., indicated ages $>$ 2500 Myr) with excesses. The following twelve stars have at least four independent age determinations (e.g., four measurements of the CAI) and should have relatively reliable ages: from DUNES,  {\it HIP 14954}, F8V, 4500 Myr; {\it HIP 15372}, F2V, 4000 Myr;  {\it HIP 32480}, G0V, 5000 Myr;  {\it HIP 65721}, G4V, 8300 Myr; {\it HIP 73100}, F7V, 4000 Myr; {\it HIP 85235}, K0V, 5600 Myr; {\it HIP107649}, G2V, 4000 Myr; and from DEBRIS,  {\it HD 7570}, F9V, 5300 Myr; {\it HD 20010}, F6V, 4800 Myr (there is only one measurement of the CAI, but multiple measurements put logg $<$ 4); {\it HD 20794}, G8V, 6200 Myr;  {\it HD 102870}, F9V, 4400 Myr; and {\it HD 115617}, G7V, 5000 Myr. The number of determinations for the following three stars are lower, but there is no credible evidence that the age estimates are seriously incorrect: {\it HIP 27887} K2.5V, DUNES, 5500 Myr; {\it HIP 71181} K3V, DUNES, 3000 Myr; and {\it HD 38393} F6V, DEBRIS, 3000 Myr.  Based on their positions on the HR diagram and a review of other data, we revised the age in \citet{gaspar2013} for {\it HIP 32480}, G0V, from 5000 to 3000 Myr and for {\it HIP 62207}, G0V, from 6400 to 1000 Myr. The CAI age of  {\it HD 61421}, F5 IV-V, from DEBRIS, 2700 Myr, is rather uncertain because it has a highly variable activity level \citep{isaacson2010}, but its position on the HR diagram is roughly in agreement with the assigned age. In summary, the great majority of the relatively old stars with excesses appear to have reliable age estimates. as was also the case for the {\it Spitzer} sample. These stars are also distributed over the full range of spectral types in the sample.

\subsection{Combined Sample}

Because of the similarity in behavior between the {\it Spitzer} and {\it Herschel} samples, we can combine them directly. We now use this combined sample to obtain a more detailed picture of the evolution of the far infrared emission by debris disks. First, we will test for the overall evolution between the incidence of excesses in young and old disks. Below, we will justify a division between a "young" sample with ages $<$ 3000 Myr and an "old" one with ages $>$ 4500 Myr, with the gap left as a transition. There are 235 stars in the first and 168 in the second category. We compute the excess probabilities based on different threshold values of R, and Table 5 shows the results. With the added statistical weight from combining the two samples, there is a well-detected drop in the incidence of excesses with age. The result is seen at all of the excess-detection thresholds (i.e., all values of R), although it becomes stronger for larger excesses. At the lowest threshold, R = 1.5, it is significant roughly at a level of 5.5 $\sigma$. 

Two thirds of this combined sample are detected at $\ge$ 3 $\sigma$ (or have photospheric levels that correspond to 3 $\sigma$ or greater), reducing the probability of censoring biases. Nonetheless, we repeated the confirmation test described in Section 4.3 for the merged sample\footnote{The different distribution of ages in the combined sample, together with the more rapid evolution for large excesses, results in a different log-normal fit.}. The model shows that the combined sample is uniform in overall sensitivity to debris disk excesses across the entire age range (e.g., fitted values of 34 - 35 excesses in each of the younger and older halves of the "young sample" and to the "old" sample).  The prediction of 35 for the old sample with a variance by the Monte Carlo simulation of 2.2 is significantly larger than the observed number of 11. The probablility of this outcome (computed from the binomial distribution) if there is no evolution is $<$ 2 $\times$ 10$^{-7}$, or equivalently it is a $>$ 5 $\sigma$ indication of evolution. 

Given the confirmation of the K-M results and the strong indication of evolution, we applied the K-M approach to estimate the incidence of debris excesses as a function of age. We have used an 80-star running average of the combined sample to determine the most probable incidence of excesses for various values of R. The results are displayed in Figure 7 and selected values are listed in Table 6. As shown by the 1-$\sigma$ error zones, the individual bumps and wiggles in the curves are not significant. The important characteristics of the behavior are: 1.) the very large excesses decline more rapidly than smaller ones; 2.) small excesses (R $\sim$ 2) persist at a more-or-less constant incidence up to an age of $\sim$ 3000 Myr; 3.) there is a decline in the incidence of excesses between 3000 and 5000 Myr; and 4.) there is a persistent (but spectral-type-independent) incidence of excesses for stars older than 5000 Myr, even though an extrapolation of the decay from 3000 to 5000 Myr would suggest they should be rare. 

Previous work has found hints of the overall reduction of excesses at old ages, but has been hampered by inadequate statistics (e.g., \citet{bryden2006, carpenter2009}, or by a number of excesses attributed to very old stars that contradicted the trend \citep{trilling2008}. The larger sample, higher quality measurements, and emphasis on accurate age determination can account for the trend seen clearly in our study whereas there were only hints in these previous ones. 

In comparing these results with theoretical predictions, we take a fiducial debris disk radius of 30 AU \citep{gaspar2013} and a typical cold disk temperature of 60K \citep{ballering2013}. From the latter, we can estimate fractional excesses corresponding to values of R as listed in Table 7. At R = 2, the constant level of excess up to a drop between 3000 and 5000 Myr is then in approximate agreement in pattern of behavior with the simple models for steady-state disk evolution in \citet{wyatt2008}, Figure 5. 
A quantitative comparison with our results, however, cannot be made with the predicted decay for a single disk; a synthesis over a population of disks is required. Figure 7 shows such a model. It is for a population of debris systems around G0V stars with an initial mass distribution width as in \citet{andrews2005} with a median mass of $5.24 \times 10^{-6}$ Earth masses (integrated up to particles of 1mm radius). The debris is placed at a distance of 32.5 AU from the stars in narrow rings (10\% width and height) and with optical and mechanical properties for icy grains. Details of the model are as in \citet{gaspar2013}. This model matches the initial relatively constant incidence of disks up to $\sim$ 3 Gyr and the decay from that age to $\sim$ 5 Gyr well. However, it falls well below the observations at ages $>$ 5 Gyr.  In general, the decay timescale is roughly inversely proportional to fractional excess, but with indications of more excesses than predicted for old stars.  It is likely that the oldest disks represent late phases of dynamical activity, as derived also by \citet{gaspar2013} with a detailed digital model of steady-state evolution that reproduces the number of excesses but yields a distribution of excess amounts that falls significantly short of the observations.

\section{Conclusions}
We present a sample of 238 F4 - K2 stars with high-quality {\it Spitzer} observations through 70 $\mu$m, and for which we have been able to determine accurate ages from the chromospheric activity index, X-ray emission, and placement on an HR-diagram. Consistent with previous work, we find that the incidence of excesses at 24 $\mu$m decays dramatically from $13.6 \pm 2.8$ \% for stars $<$ 5 Gyr in age to none for older stars. At 70 $\mu$m, we also find a significant reduction in the incidence of excesses, from 38, or $22.5 \pm 3.6$ \% to 4, or $4.7^{+3.7}_{-2.2}$ \%. We have evaluated the significance of this apparent reduction in the incidence of 70 $\mu$m excesses in two ways, both of which correctly account for the variable detection limits and resulting censoring of the data. The first is the traditional Kaplan-Meier test, which  indicates a decay by greater than 2$\sigma$ significance, nominally by a factor of 4.5. In the second method, we determined the distribution of excess ratios among the young (ages $<$ 5 Gyr) stars and assumed it could be treated as a probability distribution to determine if a specific star is likely to have an excess above its individual detection limit. This model was used to predict the number of excesses that would have been detected in the old ($>$ 5 Gyr) sample if they behaved identically to the young sample. This approach also predicts a decay at greater than 2 $\sigma$ significance, and nominally by a factor of five. The two methods therefore agree well on the existence and amount of this decay. 
Within this sample, there are two stars with excesses at 24 $\mu$m but no strong excess at 70 $\mu$m, HD 1466 and HD 13246.

We then combined the {\it Spitzer}-only stars with the measurements for the DEBRIS and DUNES programs (for the former using both {\it Spitzer} and {\it Herschel} data to improve the weight of the measurements). This combined sample is doubled in size, allowing us to reach robust conclusions about debris disk evolution in the far infrared. Within this larger sample, we examine the relative incidence of excesses around stars of types F4 - K4, finding no statistically significant trend with spectral type. We therefore analyze the evolution of the entire sample together. We find that, at fractional debris disk luminosity of just under $10^{-5}$, the incidence of far infrared excesses declines significantly between 3000 and 5000 Myr, while for fractional luminosity of just under $10^{-4}$, the decline is at an order of magnitude younger ages. This behavior is consistent with a theoretical model for passive disk evolution. However, there appear to be more excesses for the oldest stars than predicted by passive-evolution models. These excesses are distributed over the full range of spectral types in our sample. This behavior suggests that the oldest debris disks are activated by late-phase dynamical activity.

\acknowledgements
We thank the anonymous referee for a very helpful report, and the editor for suggesting improved statistical approaches. We have made use of the SIMBAD database and the VizieR catalog access tool, operated at CDS, Strasbourg, France. We have also used data products from the Two Micron All Sky Survey, which is a joint project of the University of Massachusetts and the Infrared Processing and Analysis Center/California Institute of Technology, funded by the National Aeronautics and Space Administration and the National Science Foundation. This work is also based, in part, on observations made with {\it Spitzer Space Telescope}, which is operated by the Jet Propulsion Laboratory, California Institute of Technology under NASA contract 1407. We thank Patricia Whitelock and the South African Astronomical Observatory for assistance and use of facilities to obtain near infrared photometry. This work supported by contract 1255094 from Caltech/JPL to the University of Arizona. JMS is supported by the National Science Foundation Graduate Research Fellowship Program, under grant No. 1122374.

\eject

\begin{deluxetable}{ccp{10cm}c} 
\tabletypesize{\scriptsize}
\rotate
\tablecolumns{4}
\tablewidth{0pt}
\tablecaption{Sources of Data}
\tablehead{\colhead{PID}           &
                 \colhead{number of stars}        &
                 \colhead{sample description}    &
                 \colhead{source}                  \\ }

\startdata

41     &      66     &   unbiased sample of F5-K5 stars (selection on expected photospheric brightness as well as spectral type), supplemented by stars with radial velocity planets but otherwise similar    &  \citet{trilling2008}  \\  

52     &      1     &  volume limited sample , with additional constraints on predicted flux and backgrounds    &  program abstract  \\  

54     &      1     &  A3-F8 stars in binary systems, selected for proximity, binary characteristics, low IR background    &  \citet{trilling2007}  \\  

72     &      8     &   young based on lithium, H$\alpha$, X-rays, or membership in TW Hya    &  \citet{low2005}  \\  

80     &      10     &   A through M-type dwarfs selected on the basis of probable youth from lithium abundance, X-ray emission, chromospheric activity, and rotation    &  \citet{plavchan2009}  \\  

84     &      3     &  F, G, and K stars indicated to be members of Sco/Cen based on Hipparcos    &  \citet{chen2005}  \\  

148     &     78     &   unbiased sample from 3 Myr to 3 Gyr in estimated age, equally distributed over this range, masses between 0.7 and 1.5 M$_\odot$ (roughly F2V - K5V)    &  \citet{meyer2006}  \\  

206     &     1     &   eight members of Castor Moving Group, no preselection for excesses   &  program abstract  \\
\enddata
\end{deluxetable}

\eject

\begin{deluxetable}{rcrrrrrrrrrcc} 
\tabletypesize{\scriptsize}
\rotate
\tablecolumns{13}
\tablewidth{0pt}
\tablecaption{{\it Spitzer} Star Sample}
\tablehead{\colhead{HD}                   &
           \colhead{ST\tna}               &
           \colhead{Age}                  &
           \colhead{Ref.$^b$}                 &
           \colhead{$\pi$}                &
           \colhead{[Fe/H]}               &
           \colhead{$V$}                  &
           \colhead{$K_S$}                &
          \colhead{Ref.$^c$}          &
	\colhead{PID}                  &
           \colhead{AORKey}               &
           \colhead{$F_{24}$ (error)}     &
           \colhead{$F_{70}$ (error)}     \\
	   \colhead{}                     &
	   \colhead{}                     &
	   \colhead{MYr}                  &
              \colhead{}                     &
	   \colhead{(mas)}                &
	   \colhead{(dex)}                &
	   \colhead{(mag)}                &
	   \colhead{(mag)}                &
	   \colhead{}                     &	  
 	   \colhead{}                     &
	   \colhead{}                     &
	   \colhead{(mJy)}                &
	   \colhead{(mJy)}                }
\startdata
           105 &             G0V &      170 &          1,2,3,4,5,6 &     25.4 &    -0.23 &      7.51 &      6.12 &     1&     148 &      5295872 &     27.82 (0.30) &    152.70 (9.71) \\
            377 &             G2V &      220 &            1,3,7,5,6 &     25.6 &    -0.10 &      7.59 &      6.12 &    1&      148 &      5268992 &     36.31 ( 0.39) &    170.60 (10.62) \\
            870 &             K0V &     2300 &                  2,8 &     49.5 &    -0.20 &      7.22 &      5.38 &    1&    30490 &     19827456 &     49.75 (0.51) &     22.51 (4.98) \\
            984 &              F5 &      250 &              1,3,6,7 &     21.2 &    -0.11 &      7.32 &      6.07 &   1&       148 &      5272064 &     26.51 (0.29) &     -7.78 (5.57) \\
           1237 &         G8.5Vk: &      300 &              1,2,6,8 &     57.2 &    -0.09 &      6.59 &      4.86 &    1&        41 &      4060672 &     84.41 (0.85) &     11.81 (2.08) \\
           1461	&	G0V	&	7200	& 3,5,9	& 43.0	&	0.09	&	6.47	&	4.897	&  1&	30490	&	19886336	&	80.19	(0.35)	&	61.43	(6.08)	\\
           1466 &             F9V &      200 &                  2,6 &     24.1 &    -0.27 &      7.46 &      6.15 &      1&     72 &      4539648 &     34.47 (0.36) &     16.83 (3.09) \\
	2151	&	G0V	&	7000	& 2,8,15,17 & 134.1	&	-0.03	&	2.82	&	1.286	&  2,3,4,5 &	30490	&	19759616	&	2225.00	(1.17)	&	250.70	(5.26)	\\
\enddata
\tablenotetext{a}{Spectral Type}
\tablenotetext{b}{Age references: 1 Schroeder et al. 2009; 2 Henry et al. 1996; 3 Wright et al. 2004; 4 Rocha-Pinto \& Maciel 1998; 5 Isaacson \& Fischer 2010; 6 X-ray from RASS; 7 White et al. 2007; 8 Gray et al. 2006; 9 Gray et al. 2003; 10 Martinez-Arnaiz et al. 2010; 11 Kasova \& Livshits 2011; 12 Jenkins et al. 2006; 13 Zuckerman \& Song 2004; 14 Montes et al. 2001; 15 logg < 4; 16 Duncan et al. 1991; 17 Buccino \& Mauas 2008; 18 Rizzuto et al. 2011; 19 Barrado y Navascues 1998; 20 Karatas et al. 2005; 21 Schmitt \& Liefke 2004; 22 Ng \& Bertelli 1998; 23 Mamajek 2009; 24 Mishenina et al. 2011}
\tablenotetext{c}{Sources of K photometry: 1 2MASS; 2 this work; 3 \citet{carter1990}; 4 \citet{glass1974}; 5 \citet{bouchet1991}; 6 \citet{johnson1966a}; 7 \citet{selby1988}; 8 \citet{kidger2003}; 9 \citet{alonso1998}; 10 \citet{aumann1991}; 11 \citet{allen1983}; 12 \citet{UKIRT1994}}

\end{deluxetable}

\eject

\eject

\begin{deluxetable}{lcrrr}
\tabletypesize{\footnotesize}
\tablecolumns{5}
\tablewidth{0pt}
\tablecaption{Stars with Excesses}
\tablehead{\colhead{HD}                   &
           \colhead{Excess?\tna}          &
           \colhead{$K_{S}-[24]$}         &
           \colhead{$R_{70}$}             &
           \colhead{$\chi_{70}$}          }
\startdata
105	&	2	&	0.09	&	54.47	&	24.99	\\
377	&	3	&	0.38*	&	80.87	&	26.40	\\
870	&	2	&	-0.01	&	4.10	&	3.55	\\
1461	&	2	&	0.02	&	7.10	&	8.65	\\
1466	&	3	&	0.35*	&	5.46	&	5.30	\\
6963	&	2	&	0.08	&	12.72	&	8.31	\\
7193	&	1	&	0.11	&	3.04	&	1.76	\\
7590	&	2	&	0.04	&	35.42	&	38.02	\\
\enddata
\tablenotetext{a}{1 = excess at 24 $\mu$m only (or no 70 $\mu$m measurements), 2 = excess at 70 $\mu$m only, 3 = excesses at both 24 $\mu$m and 70 $\mu$m}
\tablenotetext{*}{24 $\mu$m excess confirmed by WISE W3-W4 color.}
\end{deluxetable}

\eject


\begin{deluxetable}{crccc}
\tabletypesize{\footnotesize}
\tablecolumns{5}
\tablewidth{0pt}
\tablecaption{Probability Distribution Parameters for the {\it Spitzer} Sample}
\tablehead{\colhead{Shape of Fit}           &
           \colhead{$\chi^{2}$}             &
           \colhead{Normalization}          &
           \colhead{$\mu$\tna}              &
           \colhead{$\sigma$}               }
\startdata
Power Law & 20.6 & 0.219 & 1.50 & - \\
Lognormal & 6.2 & 0.448 & 1.50 & 2.12 \\
Exponential & 139.8 & 0.284 & 25.9 & - \\
Gaussian & 401.1 & 0.297 & 25.4 & 24 \\
\enddata
\tablenotetext{a}{$\mu$ for the power law and exponential distributions represents the exponential factor.}
\end{deluxetable}

\eject

\begin{deluxetable}{ccc}
\tabletypesize{\footnotesize}
\tablecolumns{4}
\tablewidth{0pt}
\tablecaption{Probable Incidence of Young and Old Disks for the Combined Sample}
\tablehead{\colhead{R threshold}           &
           \colhead{0 - 3000 Myr$^a$ }              &
           \colhead{$\ge$ 4500 Myr$^a$}               }
\startdata
1.5  &  0.32$^{+0.06}_{-0.06}$  & 0.08$^{+0.06}_{-0.03}$\\
2  & 0.25$^{+0.06}_{-0.05}$   & 0.05$^{+0.05}_{-0.03}$ \\
3  &  0.21$^{+0.06}_{-0.05}$  & 0.04$^{+0.05}_{-0.02}$ \\
4 &  0.18$^{+0.06}_{-0.05}$ & 0.02$^{+0.04}_{-0.01}$ \\
5 &  0.17$^{+0.055}_{-0.045}$ & 0.01$^{+0.04}_{-0.01}$ \\
10 &  0.11$^{+0.05}_{-0.04}$ & 0.00$^{+0.01}_{-0.00}$ \\
\enddata
\tablenotetext{a}{Errors are for 95\% confidence limits.}
\end{deluxetable}

\eject

\begin{deluxetable}{cccc}
\tabletypesize{\footnotesize}
\tablecolumns{4}
\tablewidth{0pt}
\tablecaption{Probable Incidence of Disks vs. Age and R threshold$^a$, for the Combined Sample}
\tablehead{\colhead{Age range (Gyr)}           &
           \colhead{Most probable fraction}          &
           \colhead{95\% confidence upper limit}              &
           \colhead{95\% confidence lower limit}               }
\startdata
280  &  0.39 .. 0.30 .. 0.19 .. 0.13  & 0.51 .. 0.41 .. 0.29 .. 0.22 & 0.29 .. 0.20 .. 0.11 .. 0.07 \\
500  & 0.30 .. 0.24 .. 0.15 .. 0.11  & 0.41 .. 0.35 .. 0.25 .. 0.21 & 0.20 .. 0.16 .. 0.09 .. 0.06 \\
850  &  0.21 .. 0.21 .. 0.19 .. 0.12 & 0.32 .. 0.32 .. 0.30 .. 0.22 & 0.13 .. 0.13 .. 0.12 .. 0.07 \\
1400 &  0.24 .. 0.21 .. 0.15 .. 0.06 & 0.35 .. 0.32 .. 0.25 .. 0.15 & 0.15 .. 0.12 .. 0.08 .. 0.02 \\
2200 &  0.32 .. 0.21 .. 0.12 .. 0.08 & 0.44 .. 0.32 .. 0.21 .. 0.16 & 0.22 .. 0.13 .. 0.06 .. 0.03 \\
3300 &  0.27 .. 0.16 .. 0.08 .. 0.06 & 0.38 .. 0.26 .. 0.16 .. 0.15 & 0.18 .. 0.09 .. 0.04 .. 0.02 \\
4450 &  0.12 .. 0.08 .. 0.04 .. 0.03 & 0.22 .. 0.17 .. 0.11 .. 0.10 & 0.06 .. 0.035 .. 0.01 .. 0.00 \\
5400 &  0.08 .. 0.04 .. 0.01 .. 0.00 & 0.17 .. 0.12 .. 0.08 .. 0.06 & 0.03 .. 0.01 .. 0.00 .. 0.00 \\
6300 &  0.09 .. 0.07 .. 0.03 .. 0.00 & 0.19 .. 0.16 .. 0.10 .. 0.06 & 0.04 .. 0.03 .. 0.01 .. 0.00 \\
7500 &  0.08 .. 0.04 .. 0.01 .. 0.00 & 0.17 .. 0.12 .. 0.07 .. 0.06 & 0.04 .. 0.01 .. 0.00 .. 0.00\\

\enddata
\tablenotetext{a}{The successive entries in each cell are for R $\ge$ 1.5, 2, 5, \& 10.}
\end{deluxetable}

\eject

\begin{deluxetable}{cc}
\tabletypesize{\footnotesize}
\tablecolumns{4}
\tablewidth{0pt}
\tablecaption{Approximate Fractional Excesses vs. R}
\tablehead{\colhead{R at 85$\mu$m}           &
           \colhead{Fractional Excess}   }
\startdata
1.5  & $4 \times 10^{-6}$   \\
2  &  $9\times 10^{-6}$   \\
3  &  $1.7 \times 10^{-5}$ \\
4 &  $2.5 \times 10^{-5}$  \\
5 &    $3.5 \times 10^{-5}$ \\
10 &  $8 \times 10^{-5}$  \\
15 &  $1.2 \times 10^{-4}$  \\
20 &  $1.6 \times 10^{-4}$  \\
25 &  $2 \times 10^{-4}$  \\

\enddata
\end{deluxetable}
\eject
\eject
\eject

\newpage

\eject

\begin{figure*}
\figurenum{1}
\label{fig1}
\plotone{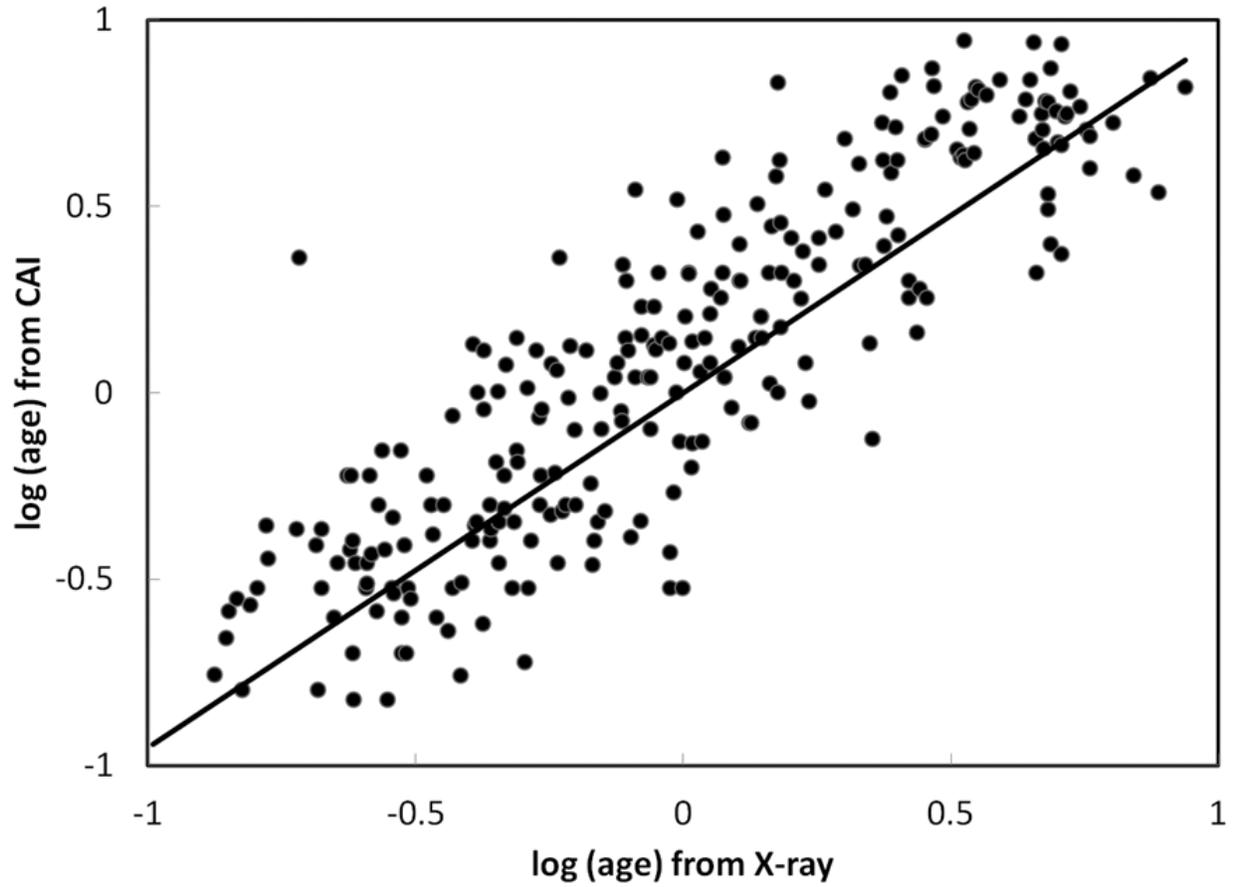}
\caption{Comparison of ages (in Gyr) from the CAI with those from X-ray luminosity. The fitted line has a slope of 1.28.} 
\end{figure*}

\eject

\begin{figure*}
\figurenum{2}
\label{fig2}
\plotone{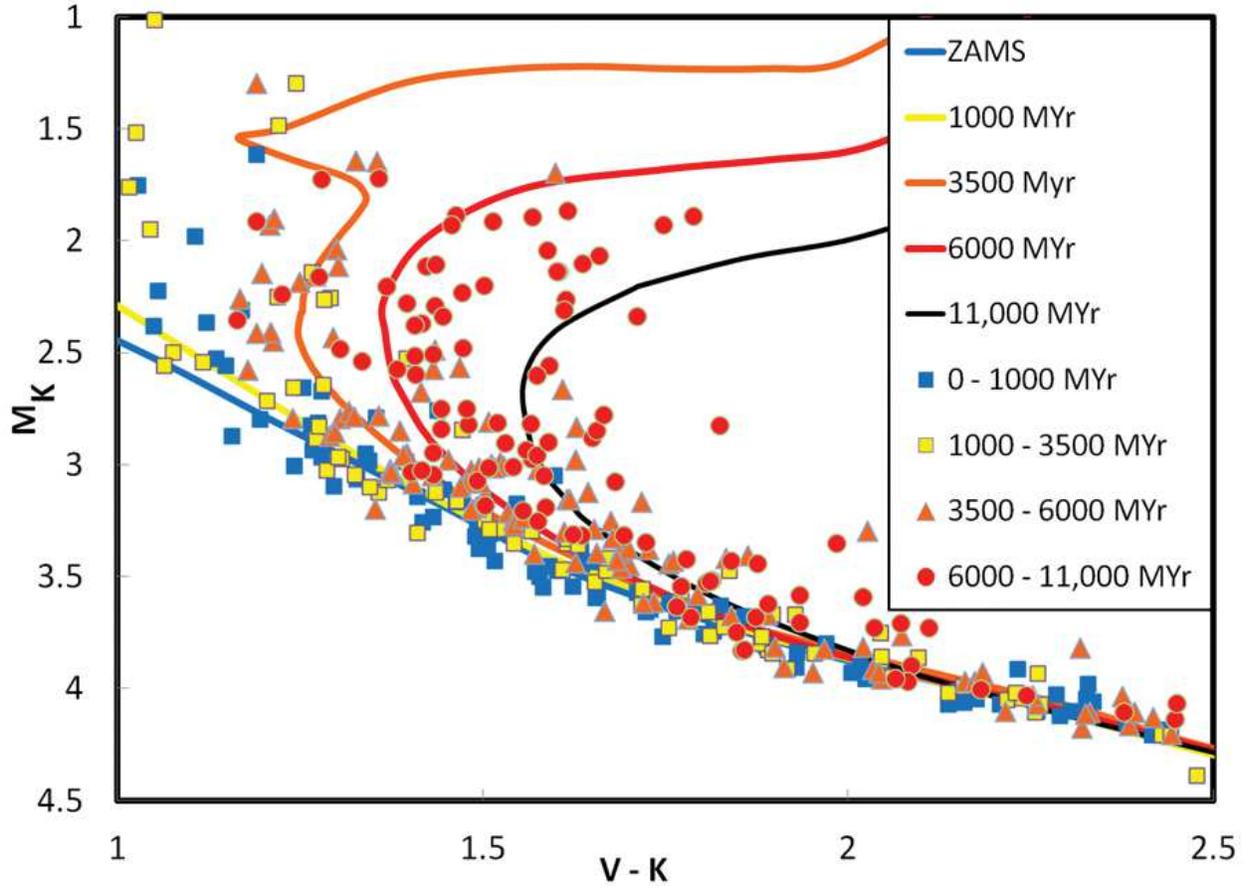}
\caption{Hertzsprung-Russell diagram prepared as described in the text to compensate for stellar metallicity and to match the observational photometric system. Ages determined from the CAI are coded by the symbols and can be compared with the isochrones for various ages \citep{bonatto2004, marigo2008}.} 
\end{figure*}

\eject

\begin{figure*}
\figurenum{3}
\label{fig3}
\plotone{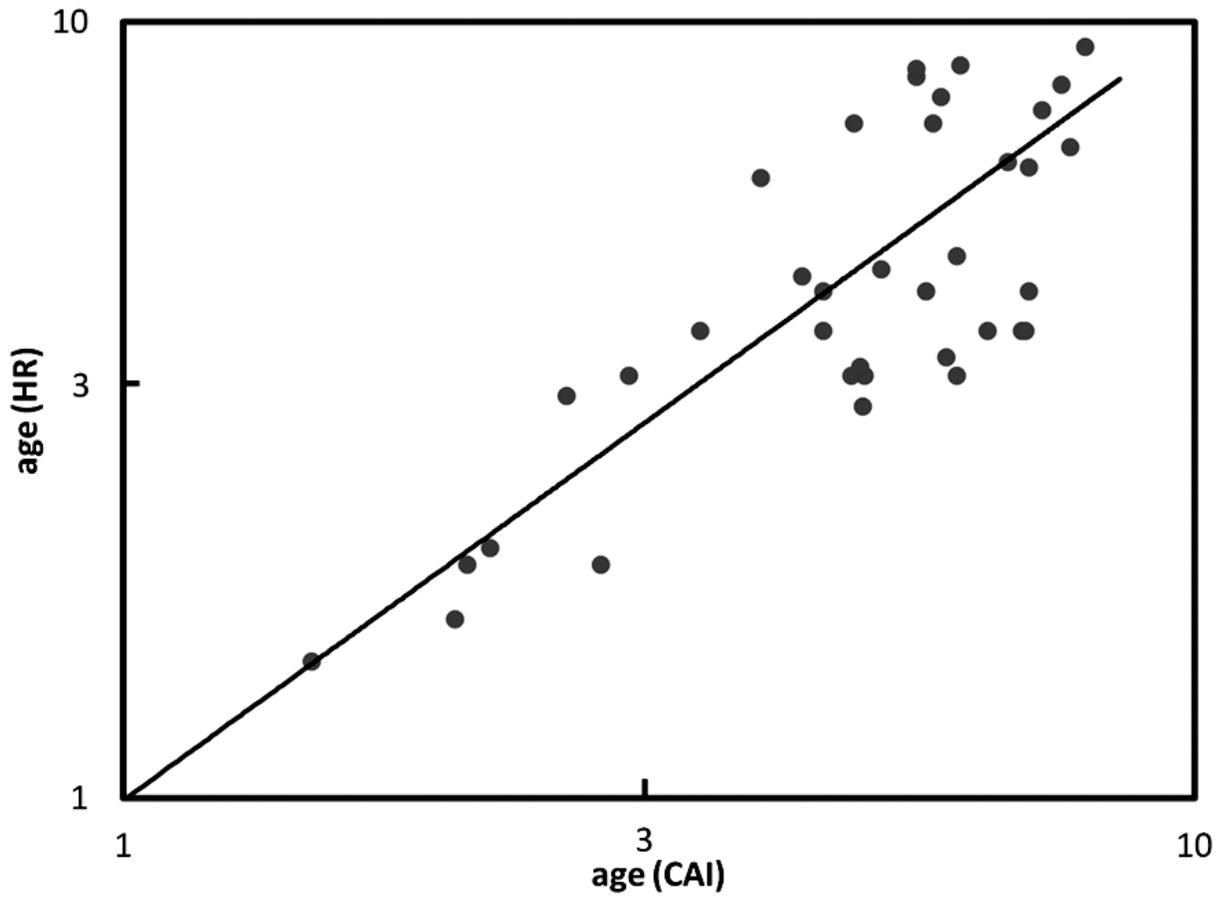}
\caption{Comparison of ages (in Gyr) from the CAI and the HR diagram. The fitted line has a slope of 0.985.} 
\end{figure*}

\eject

\begin{figure*}
\figurenum{4}
\label{fig4}
\plotone{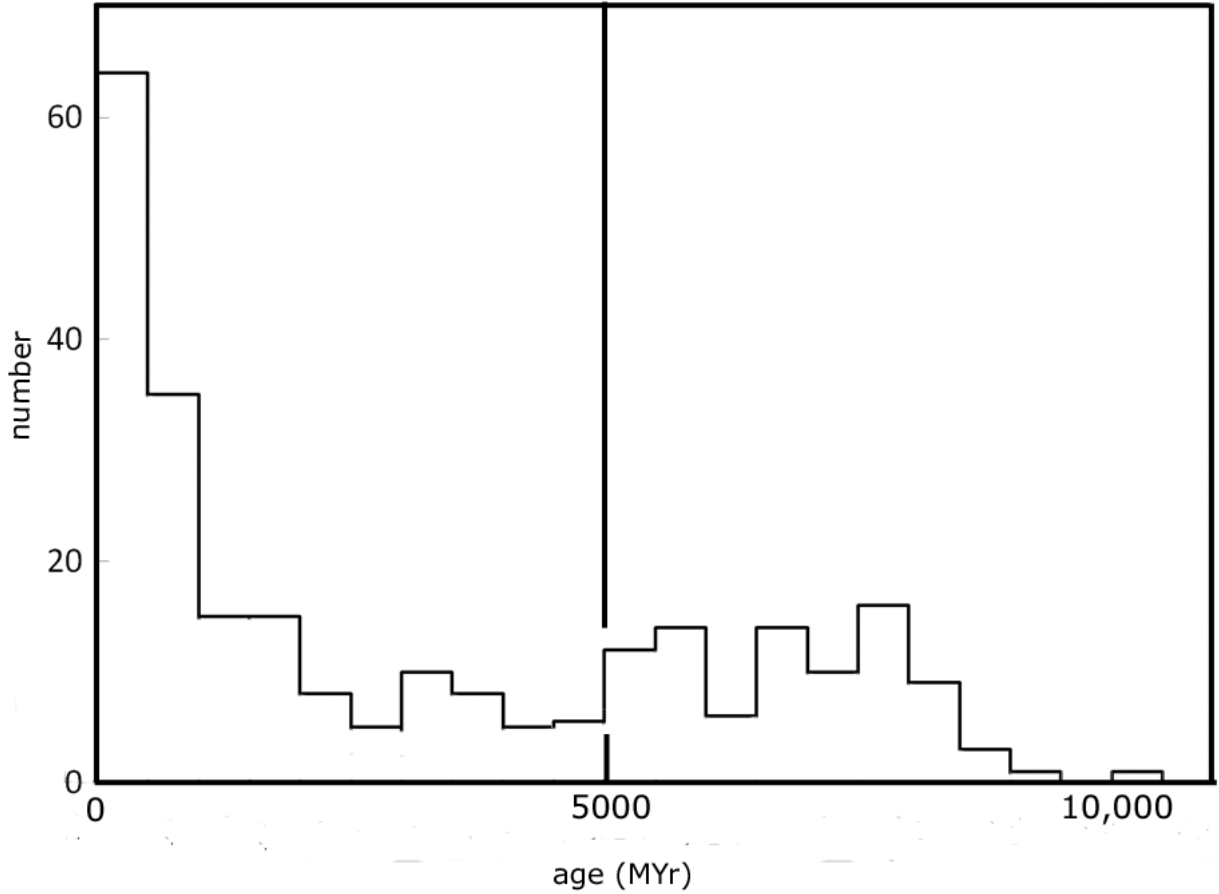}
\caption{Histogram of ages in the final samples, with a bin size of 500 Myr. The vertical line indicates the minimum age for which we could apply the HR diagram criterion to confirm the stellar age. All stars to the right of this line both have ages $>$ 5 Gyr indicated by CAI, X-ray output, $logg$ or gyrochronology, and also fall in the appropriate region of the HR diagram.} 
\end{figure*}

\eject

\begin{figure*}
\figurenum{5}
\label{fig5}
\plotone{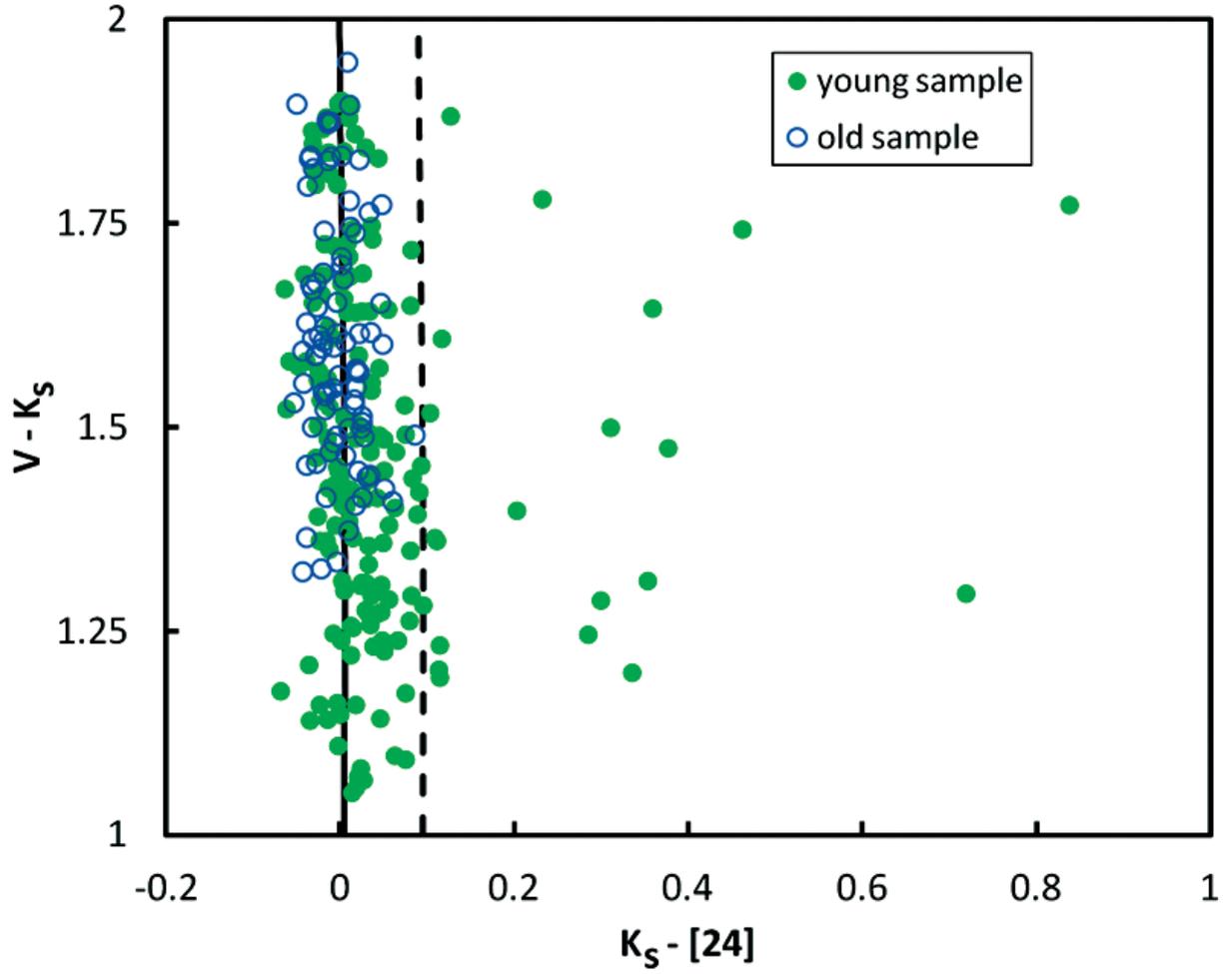}
\caption{$V-K_{S}$ vs. $K_{S}-[24]$ for the final samples. The solid line represents the zero excess average (see text), while the dashed line represents the minimum for stars with excesses. Stars with extreme excesses at 24 $\mu$m are not shown.}
\end{figure*}

\eject

\begin{figure*}
\figurenum{6}
\label{fig6}
\plotone{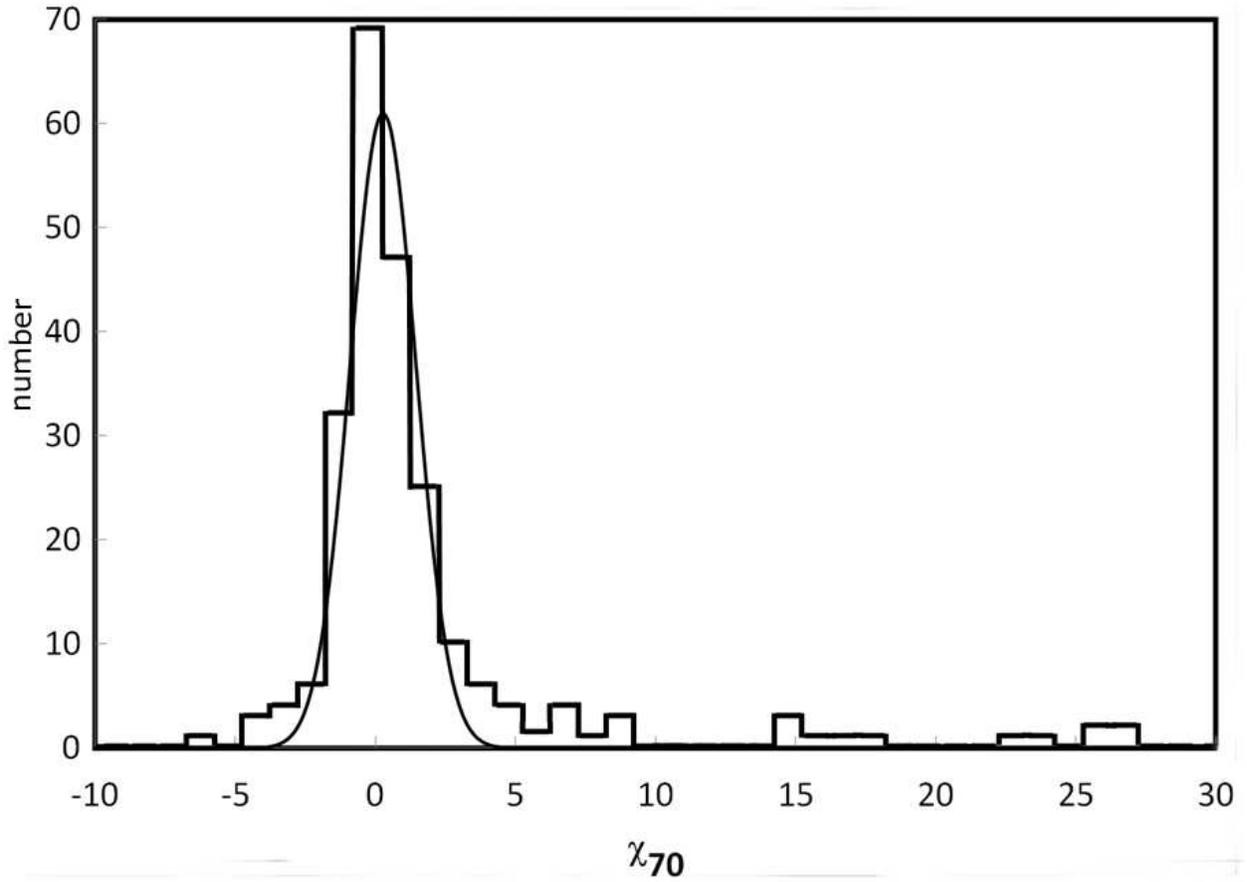}
\caption{Histogram showing the distribution of $\chi_{70}$ for our sample. Some extreme values are not shown. The Gaussian fit to the distribution indicates the behavior of the purely photospheric detections. }
\end{figure*}

\eject

\begin{figure*}
\figurenum{7}
\label{fig7}
\plotone{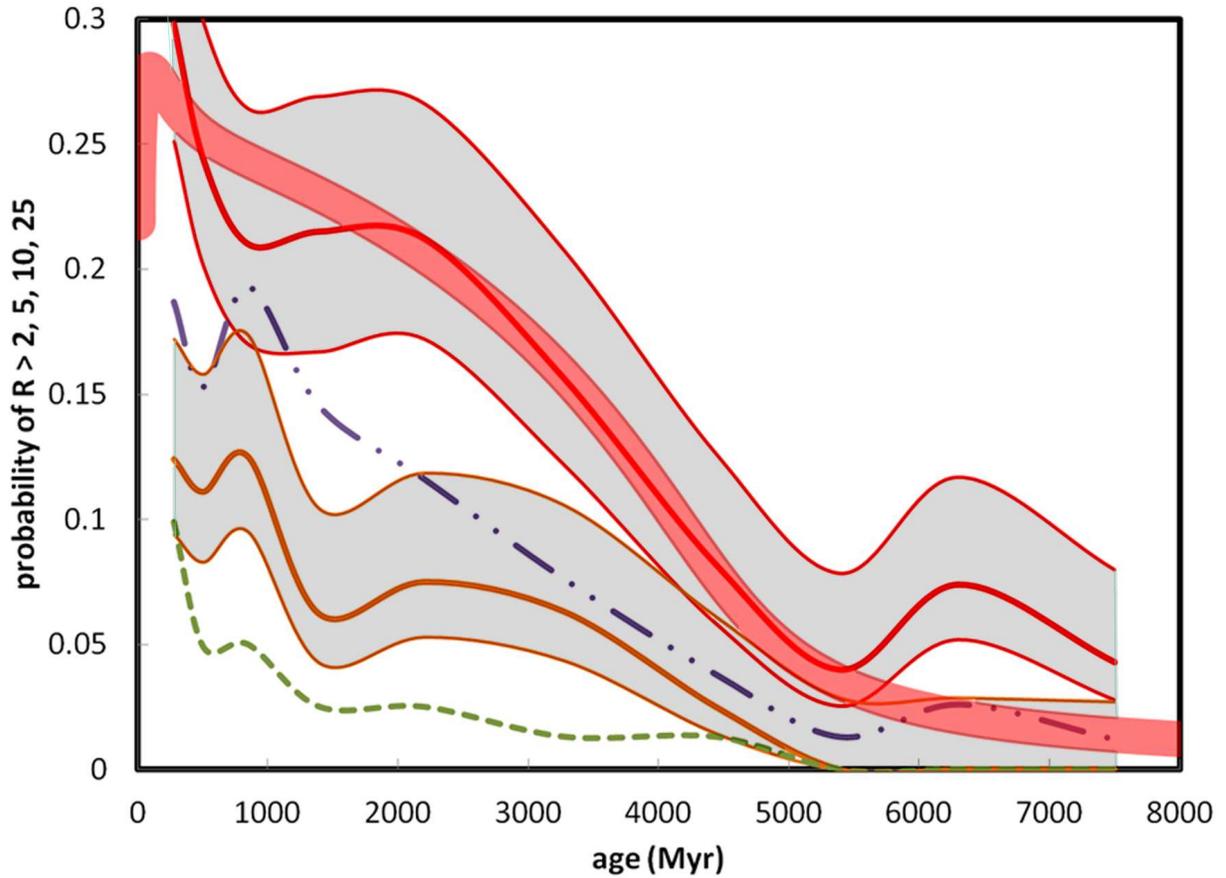}
\caption{Trend of far infrared excesses with age (as measured by R, the ratio of the measured far infrared flux density to the expected value for the stellar photosphere).
The trends are smoothed with an eighty-star running average; for R = 2 (red line) and R = 10 (orange), they are shown with grey zones indicating the 1 $\sigma$ uncertainties. The trends for R = 5 
(blue dot-dashed line) and R = 25 (green dashed line) are also shown. A theoretical model for the R = 2 case (discussed in the text) is shown as the broad red line. }
\end{figure*}


\end{document}